\documentclass[10pt]{elsarticle}

\usepackage{amssymb}

\journal{Physica A}

\begin{document}

\begin{frontmatter}



\title{Brownian dynamics study of driven partially pinned solid in the presence of square array of pinning centers:
Enhanced pinning close to the melting transition}


\author{Toby Joseph}

\address{BITS-Pilani, Dept. of Physics, K.K. Birla Goa Campus, GOA - 403726, India}

\begin{abstract}
A set of interacting vortices in $2D$ in the presence of a substrate with square symmetry and at filling ratio $1$ can display a low
temperature solid phase where only one of the reciprocal lattice vectors of the substrate is present\cite{toby1,fasano}. 
This partially pinned vortex lattice melts to a modulated liquid via a continuous transition \cite{toby1}. Brownian dynamics simulation is carried out to study the 
behavior of driven partially pinned solid at different temperatures. The average vortex velocity for forces above the depinning threshold 
shows a non-monotonic behavior with temperature, with a minimum in the average velocity close to the melting point. 
This is reminiscent of the peak effect seen in vortex systems with random disorder. This effect in the current system can be qualitatively explained 
by an effective increase in the barriers encountered by the particles as $T_c$ is approached. Approximate calculation of the energy barriers as a 
function of temperature, from the simulation supports this claim.
\end{abstract}


\end{frontmatter}


\section{Introduction}
\label{intro}
The behavior of a collection of particles in two dimensions in the presence of a periodic substrate potential has been of interest in the context 
of various types of systems. These include vortices in superconductors\cite{fasano,blatter,harada,yoav,sadovskyy,reichhardt1,reichhardt2}, modeling 
of friction with solid on solid models \cite{vanossi,granato1,song,norell}, and also experimental set ups involving colloidal aggregates in the presence of a 
substrate potential created by interfering laser beams\cite{bohlein,brazda,cao}. The periodic substrate potential could differ in symmetry, strength of the 
pinning potential (in relation to the inter-particle interaction), range of the pinning potential and density of pins. The static and dynamic behavior of the 
system varies considerably depending on these parameters \cite{toby1,harada,reichhardt3,toby2,achim}.

Driven lattice in the presence of a substrate, both random and periodic, have received a lot of attention \cite{reichhardt4} in both theoretical
and experimental studies in many of the systems mentioned above. Some of the important properties of interest include: (i) the depinning threshold force and 
the nature of depinning, \cite{achim,onuttom}, (ii) dynamical phases and transitions between them for various drives \cite{reichhardt3}and
(iii) how the properties of the system changes with system parameters like temperature and/or density \cite{bhattacharya}. The depinning phenomena
is of great relevance in the study of vortex matter, where a larger depinning threshold ensures zero resistance transport for larger applied currents.
Depinning is also crucial in the study of friction models and transport in charge density wave systems. More recently, directional mode locking behavior
has been seen in colloidal aggregates driven across periodic substrates \cite{cao}. One of the intriguing phenomena that one
observes in vortex system in superconductors with random disorder is the {\it peak effect}, wherein there is a sudden increase in the depinning force
just before an order to disorder transition \cite{bhattacharya,cha,granato2,ling,reichhardt5}.  

\begin{figure}[htbp]
\center
\includegraphics[height=8.0cm]{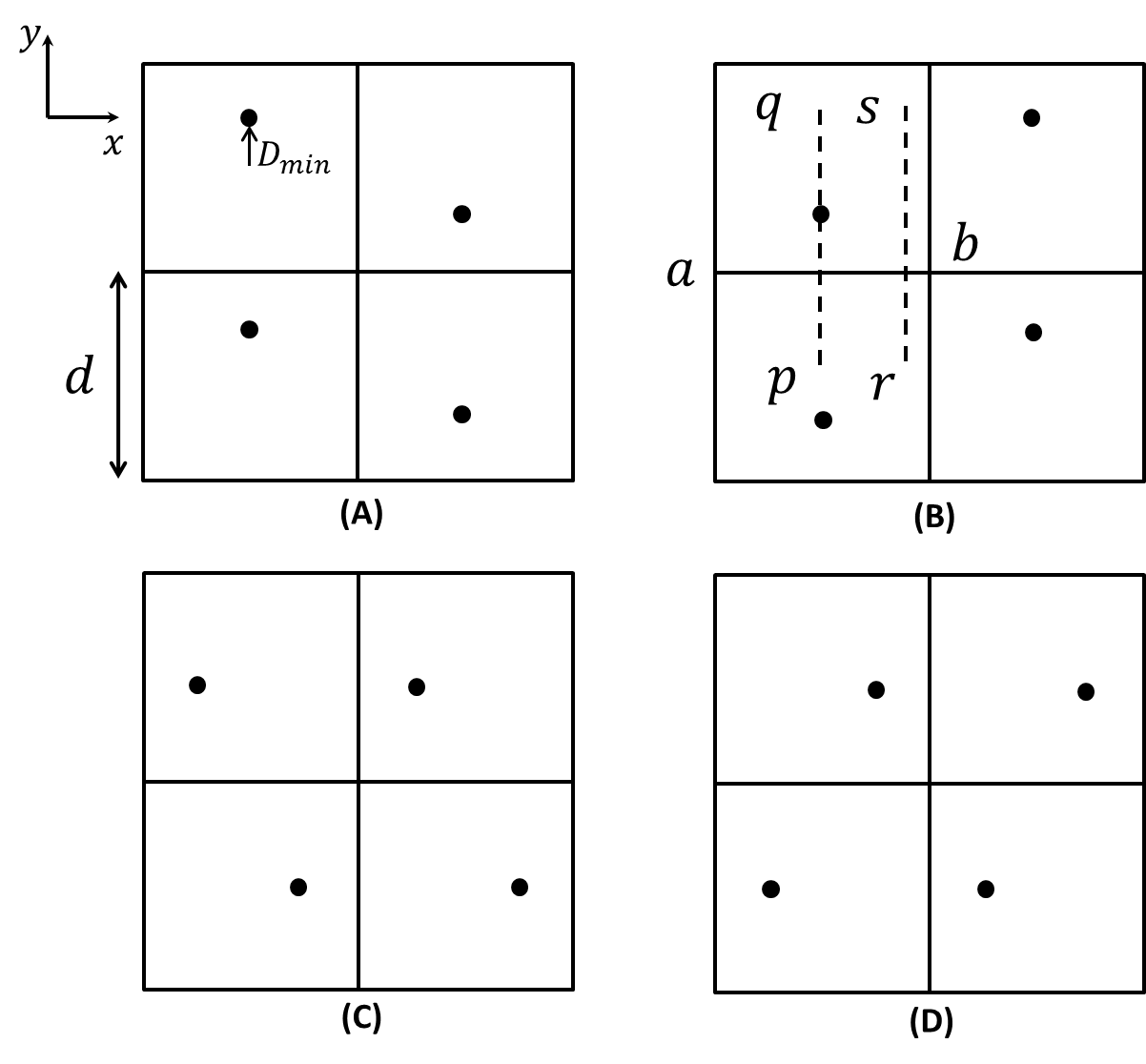}
\vspace{0.0cm}
\caption{\small The four equivalent low energy structures. The squares are drawn for visual guidance and represent the basic pinning 
squares, with the minima of the pinning potential at the centers of the squares. The unit cells for the the cases (A) and (B) are the two horizontal adjacent 
squares and for (C) and (D) are the two vertical adjacent squares. The value of  the displacement of each vortex from the center of the pinning square 
$D_{min}$, varies with the strength of the pinning potential. The lines $pq$, $rs$ and $ab$ in (B) are relevant for the discussion presented in section \ref{results}
on enhancement of resistance to vortex motion (see text for explanation).}\label{Fig01}
\end{figure}

The existence of a partially pinned solid phase of vortices in the presence of a square substrate potential (that is, substrate potential with square symmetry) was 
seen in both Monte Carlo (MC) simulations of vortex systems \cite{toby1} and in experiments \cite{fasano}, when the filling factor (which is number of vortices 
divided by number of pinning centers) is one. Similar structures have also been seen in Monte Carlo computer simulations of hard discs in square substrate 
potential \cite{neuhaus1}(referred to as the rhombic phase in the paper) and the crystal growth of such structures have been studied using density functional 
theory\cite{neuhaus2}. The partially pinned phase exhibits only one of the basic periodicity (in $x$ and $y$ directions) of the underlying 
pinning potential and is four fold degenerate. The four equivalent ground states are shown in Fig.~\ref{Fig01}. The two smallest reciprocal lattice vectors present 
in the structure shown in Fig.~\ref{Fig01} (A) are $\vec G_1 = \frac{2 \pi}{d} \hat x$ and $\vec G_2 = \frac{\pi}{d} \hat x + \frac{2 \pi}{d} \hat y$, where 
$d$ is the size of the square substrate potential unit cell. The lattice does not have a periodicity corresponding to $\vec G =  \frac{2 \pi}{d} \hat y$ which is 
present in the substrate potential and hence we refer to this structure as partially pinned solid.

In the simulation \cite{toby1} pinning centers that produce a repulsive potential with a range comparable to that of the inter-vortex interaction was considered. 
Pinning centers with these properties may be experimentally realized in vortex systems by arrays of magnetic dots each of which produces a potential that can 
be tailored \cite{pokrovsky} by adjusting its magnetic moment. Another physical realization is obtained in a square array of strong, attractive, short-range pinning 
centers at filling $n = 2$. In this case, each pinning center would trap a vortex at low temperatures, and these pinned vortices would interact with the remaining 
interstitial vortices (assuming each pin can trap only one vortex) via an effective repulsive potential \cite{shaprio}. The net potential produced by an array of such 
pinning centers has very flat minima at the centers of the square unit cells of the pin lattice. In the experiment \cite{fasano}, the vortex structure at one 
surface of a thin superconducting film with a commensurate square array of pinning centers on the other surface was imaged. The pinning of the vortex lines at one 
end on the square array provides, via their elastic energy, an effective pinning potential of square symmetry at the other surface. The interplay of this effective 
pinning and the vortex-vortex interaction at the free end leads to the partially pinned structure. A numerical modeling \cite{cornaglia} also leads to similar structure. 
Further, the MC simulations predicted a continuous melting of the partially pinned solid to a weakly modulated liquid as the temperature is increased \cite{toby1}. 
It has also seen that the form of the interaction potential (both the between the interstitial particles and that between the pinned and the interstitial particles) is not 
crucial for the behavior seen.  A model of colloidal particles interacting via screened coulomb potential also gives similar states and transitions \cite{george}. 
What seems to be relevant is that the scale of inter-particle interaction energy be comparable to that of the substrate potential.

In the current work we study the behavior of the driven partially pinned two dimensional vortex lattice. Brownian dynamics \cite{allen} simulation is carried out to
look at the behavior of the vortex lattice as it is driven along the direction parallel to the edge of the square substrate. In particular, the variation 
of particle drift as a function of temperature is studied. The structure of the vortex lattice is simultaneously monitored to correlate the 
characteristic of the drift behavior with the structural changes. The observation of a non monotonic change in vortex drift is explained using the idea of
increase in effective barrier heights close to the transition temperature. Drive along the diagonal to the square substrate has also been carried
out to check for the generic nature of the phenomenon observed.  

\section{Model and simulation}
\label{model}
The equation of motion for a Brownian particle that is obtained after the fast degrees of freedom have been projected out is a Langevin equation,
\begin{equation}
m \frac{d{\bf v}_{\alpha}}{dt} = {\bf F}_{\alpha}-\eta {\bf v}_{\alpha}+{\bf f}_
{\alpha}(t) \;\;.
\label{langevin}
\end{equation}
Here ${\bf F}_{\alpha}$ is the net force on the particle labeled by $\alpha$ (that is, the effects on the particle in question due to all kinds of interactions other than that 
with the fluid particles), $\eta$ is the friction coefficient and ${\bf f_{\alpha}}(t)$ is the noise term. The condition that at long times the particles should equilibriate to
the surrounding fluid temperature gives the form of the noise correlations,
\begin{equation}
\langle f_{i\alpha}(t)f_{j\beta}(t') \rangle = 2 \eta k_BT \delta_{ij} \delta_
{\alpha\beta} \delta(t - t').
\label{fdt}
\end{equation}
If one is just interested in the long time configurational dynamics then the inertial term $m \frac{d{\bf v}_{\alpha}}{dt}$ can be neglected (the overdamped limit)
and the equations of motion reduces to
\begin{equation}
{\bf v}_{\alpha}  = \frac{{\bf F}_{\alpha}}{\eta}+\frac{{\bf f}_{\alpha}(t)}{\eta} 
\;\;.  \label{noinertial}
\end{equation}
In order to integrate these set of equations one can discretize them~\cite{ermak} leading to the set of coupled equations,
\begin{equation}
{\bf r}_{\alpha}(t + \delta t) = {\bf r}_{\alpha}(t) + \frac{{\bf F}_{\alpha}}{\eta}
\delta t +{{\bf f}_{\alpha}}^{G}\;\;
\label{discrete}
\end{equation}
Here the components of ${{\bf f}_{\alpha}}^{G}$ are taken from a Gaussian distribution with zero average value and variance equals 
$\frac{2k_{B}T}{\eta}\delta t$. This ensures the that noise correlation has the required form as given in Eq.(~\ref{fdt}).

The force term on the right-hand side of equation (\ref{noinertial}) is a combination of three terms:
\begin{equation}
{\bf F}_{\alpha} = {\bf F}_{\alpha i} + {\bf F}_{\alpha s} + {\bf F}_{d} \;\;.
\end{equation}
Here ${\bf F}_{\alpha i} = \sum_{\beta \ne \alpha}F_0 K_1(\frac{r_{\alpha \beta}}{\lambda_{eff}})$ is the 
force on the $\alpha$th vortex from the other interstitial vortices, with $F_0$ given by $\frac{\phi_0^2}{8 \pi^2 \lambda_{eff}^3}$. 
In this expression $\phi_0$ is the flux quantum, $\lambda_{eff}$ is effective penetration depth in the thin film and 
$K_1$ is the modified Bessel function of the second kind. We approximate $\lambda_{eff} = \frac{\lambda^2}{h}$, where $\lambda$ is the penetration
depth and $h$ the film thickness. The sum is over all the other vortices and $r_{\alpha \beta}$ is the inter vortex separation. 
${\bf F}_{\alpha s}$ is the force on the $\alpha$th vortex due to the substrate potential. The substrate potential having square symmetry is 
generated by the interaction of the interstitial vortices with pinned vortices (that form a square array of lattice spacing, $d$). 
The interaction between the interstitial vortex and the pinned vortices is assumed to be of the form ${\bf F}_{\alpha s} = A F_m 
\sum_{\gamma}K_1(\frac{r_{\alpha \gamma}}{\lambda})$, where the sum is over the pinning centers and $A$ is a control parameter \cite{toby1}.
${\bf F}_{d}$ represents the driving force on the vortex due to an applied current in the perpendicular direction.
All the simulations are done for filling fraction equal to one. That is, the number of interstitial vortices is one per pinning site.
The studies have been done for various values of the substrate strength, but the focus is mainly on the regime where we have the partially 
pinned solid at low temperatures. 

The number of interstitial vortices simulated using Brownian Dynamics simulations was $N = 400$. The averages of various measured quantities were 
done over $5.0 \times 10^{5}$ time steps after letting the system evolve for about $5.0 \times 10^{4}$ time steps. The value
of $\delta t/\eta = 0.001 d/F_a$ where $F_a = F_0 K_1(10)$ is the force between two interstitial vortices separated by a distance $d$, the lattice spacing.
We have checked that this value of $\delta t/\eta$ is small enough to rule out step size dependence in the result for the temperature range we
are working with. The force was measured in units $F_{a}$, and the distance in units of the penetration depth, $\lambda$. The system parameter we used are appropriate for
Nb sample studied earlier \cite{harada}: $\lambda = 0.1 \mu m$, $d = 10\lambda$ and $h = \lambda$.  The bulk of the results we report here is for the value of substrate strength 
$A = 0.0001$ as in the case of equilibrium studies \cite{toby1}. We give below few of the important energy and force scales in the systems. We use these
as units in presenting the simulation data. The temperature equivalent of the inter-vortex interaction energy ($U_0 = F_0 \lambda_{eff} K_0(10)$)for two vortices 
separated by distance $d$ is $T_0 = \frac{F_0 \lambda_{eff} K_0(10)}{K_B}$, where $K_0$ is the modified Bessel function of second kind and $K_B$ is the
Boltzmann constant. Temperature in the plots below is given in term of $T_0$. The force that a pinned vortex exerts on a interstitial vortex at the center 
of one of the pinning squares that is nearest to it is, $F_b = A F_0 K_1(5 \sqrt{2})$. This would be the appropriate force to compare the depinning forces with.
The minimum substrate potential energy barrier that the interstitial vortex has to cross to go from center of one of the pinning square to the neighboring one is
$U_b \approx A F_0 \lambda_{eff}(2 K_0(5) - 4 K_0(5 \sqrt{2}))$. The temperature equivalent of this energy in units of $T_0$ is about $0.03$. 

\section{Results and discussion}
\label{results}
\begin{figure}[htbp]
\center
\includegraphics[height=7.0cm]{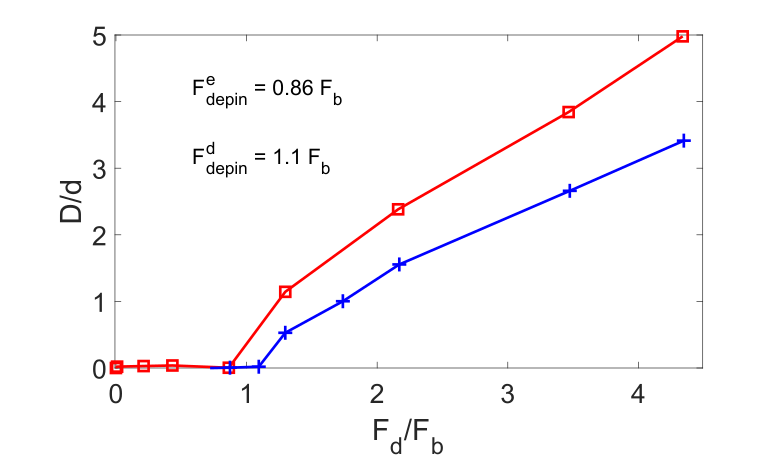}
\vspace{0.0cm}
\caption{\small Average vortex displacement $D$ (in units of lattice spacing, $d$) as a function of drive force in the easy direction (square) and 
diagonal direction (plus) at $T = 0.001$ (in units of $T_0$). The depinning force is found to be, $F^{e}_{depin} = 0.86 F_b$ 
for easy direction drive and marginally large at $F^{d}_{depin} = 1.1 F_b$ for the diagonal drive.} 
\label{Fig02}
\end{figure}
The barriers encountered when the lattice is driven along the direction of vortex displacement from the center in the ground state 
($\hat y$ direction in Fig~\ref{Fig01}(A)) is much lower than that in the perpendicular direction ($\hat x$ direction in Fig~\ref{Fig01}(A)). 
This difference stems from the asymmetry of the ground state. We refer to the easy to drive direction as the {\it easy direction}. Large drives 
lead to averaging of the underlying substrate potential along the direction of drive. That is, in the frame of the moving lattice the substrate
potential will be a drifting entity and the time average potential profile will differ for the static case. Unlike in the case of a random substrate potential, 
here one ends up with a background modulated potential whose form would depend on the direction of the drive. Drive along the easy direction 
leads to a washboard potential which has modulations of period $d$ in the direction perpendicular to the drive. The ground state of the vortex lattice 
in such an averaged out substrate is a state similar to the partially pinned structure with the displacement of the vortices from the center equal to $d/4$. 
Thus for small $A$ where this displacement is already close to $d/4$ there is not much of a structural change with drive. But for larger $A$ 
values of ($A \gtrsim 0.01$) for which the ground state of the undriven lattice has square symmetry \cite{toby1}, the lattice is modified 
appreciably by a large drive. We first look at the depinning of the vortex lattice at low temperatures for a few driving force directions.

\subsection{Zero temperature depinning}
The vortex current, $I_v$, measures the resistance to vortex motion under the action of the drive. The average vortex displacement, $D$, in 
the direction of the drive during the simulation is proportional to the average velocity of the vortices. For a given choice of 
$\frac{\delta t}{\eta}$ and fixed number of Brownian Dynamics moves one can use $D$ to compare the vortex currents. 
In all the results presented here, the time for which the lattice is driven is the same and hence $D$ has been used as a measure of the 
average vortex current. In Fig.~\ref{Fig02} the average displacement of the vortices as a function of driving force for easy direction drive as well
as for drive along the diagonal direction (($\hat y + \hat x$ direction in Fig~\ref{Fig01}(A)) at $T = 0.001$ (in units of $T_0$) is shown for the 
case $A = 0.0001$. The depinning force, determined by the onset of vortex motion, is found to be $F^{e}_{depin} = 0.86 F_{b}$ for the drive in 
the easy direction and $F^{d}_{depin} = 1.1 F_{b}$ for the drive in diagonal direction. The larger depinning force for drive along the diagonal 
direction is to be expected as the lattice will have to cross larger barriers from substrate potential as it tries to move diagonally. 
The convex shape of the depinning curve is indicative of the elastic nature of the depinning transition \cite{reichhardt2}. The depinning force for 
drive in a direction perpendicular to the easy direction is found to be $F^{p}_{depin} = 31 F_b$ and is much larger than the value for the easy direction. 
This difference is indicative of the lack of square symmetry in the partially pinned structure and could be one of the ways in which existence of a partially 
pinned solid phase can be experimentally checked.
\begin{figure}[htbp]
\center
\includegraphics[height=7cm]{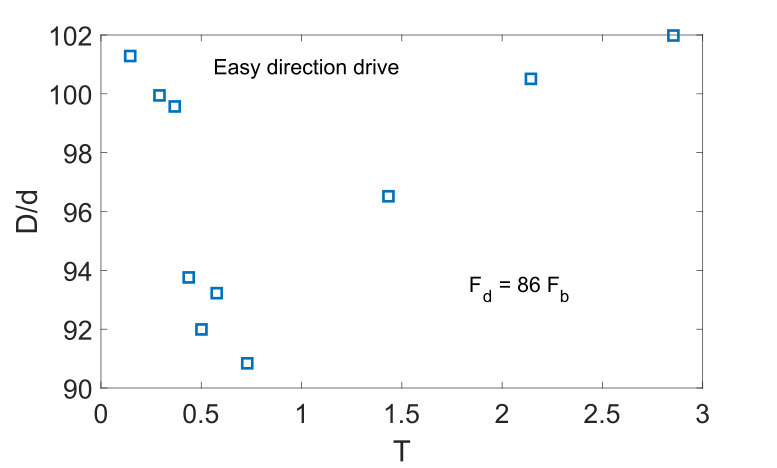}
\vspace{0.0cm}
\caption{\small  Plot showing variations of average particle velocity with temperature for driving force along the easy direction. The value of the 
substrate strength parameter, $A = 0.0001$. The average displacement of the vortices, $D$ (in unit of lattice spacing, $d$), for a fixed time 
($5 \times 10^5$ time steps) is plotted as a measure of the average vortex velocity. The driving force applied is $F_d = 86 F_{b}$. There is a 
minimum located at the vicinity of the melting transition to the modulated liquid from the partially pinned solid phase (see Fig. \ref{Fig04}). Note that
temperature is in units of $T_0$}
\label{Fig03}
\end{figure}

\subsection{Driven lattice along the easy direction}
Next we look at variation $I_v$ as a function of temperature for the case of drive in the easy direction. Keeping the driving force
value at $F_d  = 86 F_b$, which is much above the depinning threshold value for easy direction drive, average drift of the vortex lattice 
was measured for various temperature values. The vortex drift shows a non-monotonic behavior (Fig.~\ref{Fig03}) as the temperature is increased with a 
sharp fall at about $T = 0.4$ and a minima close to $T = 0.7$. In fact, the resistance to vortex lattice motion steadily increases as the temperature is increased 
from low values and reaches a maximum at around $T = 0.7$. Further increase of temperature leads to larger drift indicating easier vortex response to the drive.

\begin{figure}[htbp]
\center
\includegraphics[height=7cm]{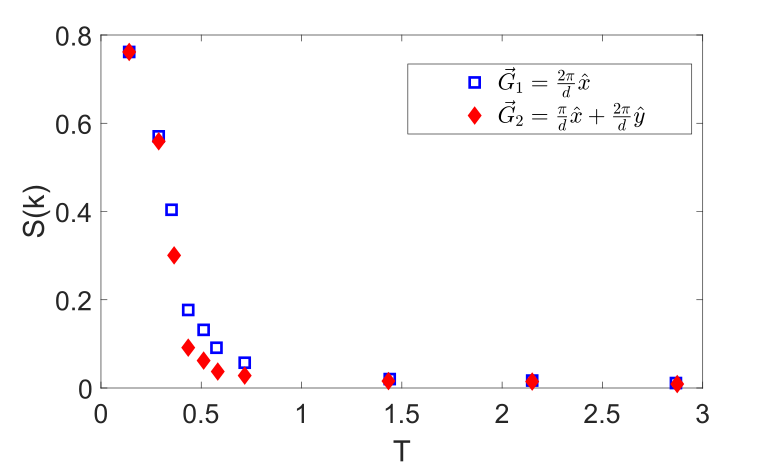}
\vspace{0.0cm}
\caption{\small  Structure factor at the two smallest reciprocal lattice vectors, $\vec G_1$ and $\vec G_2$ as a function of temperature. The value of the 
substrate strength parameter, $A = 0.0001$. The driving force applied is $F_d = 86 F_{b}$ the same as the one used for
finding the average vortex velocity (see Fig.~\ref{Fig03}). There is a transition to modulated liquid at temperature close to $T_c = 0.7$. Temperature
is given in units of $T_0$, the temperature equivalent of interaction energy between two interstitial vortices separated by substrate lattice spacing $d$.}
\label{Fig04}
\end{figure}

\begin{figure}[htbp]
\center
\begin{tabular}{cc}
\includegraphics[width=60mm]{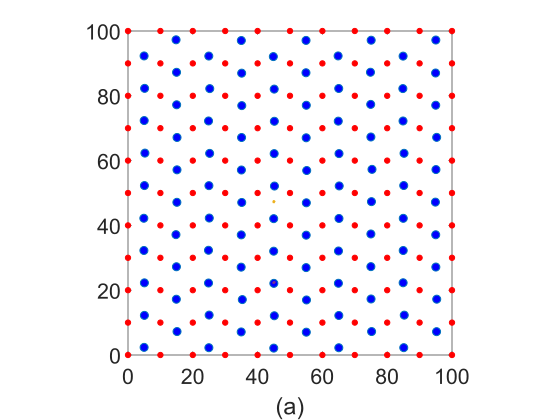}&
\includegraphics[width=60mm]{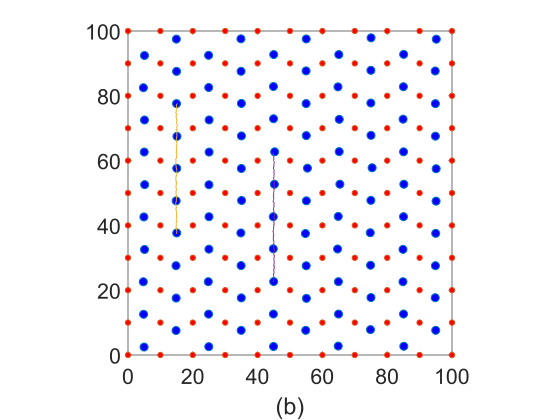}\\
\includegraphics[width=60mm]{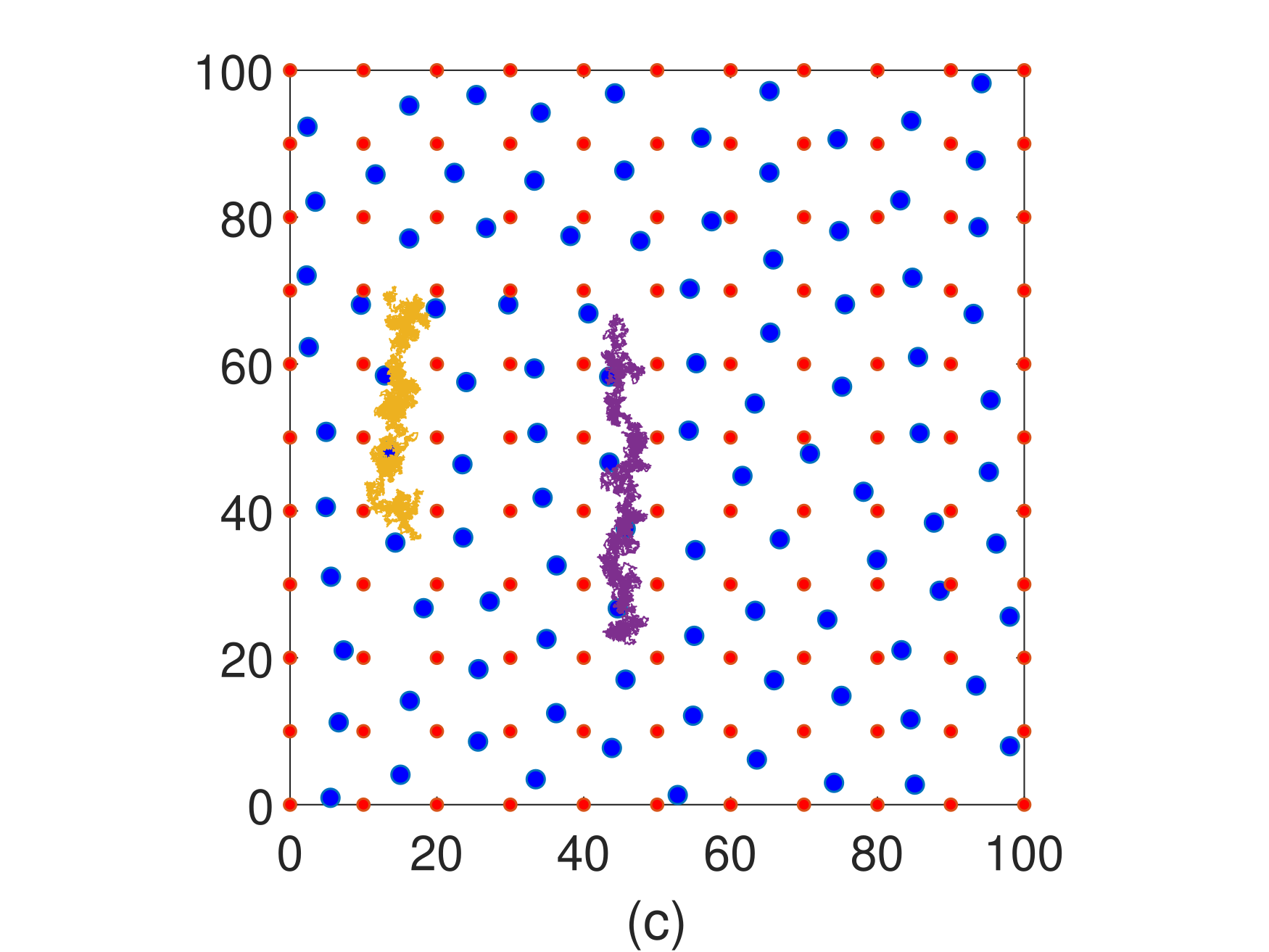}&
\includegraphics[width=60mm]{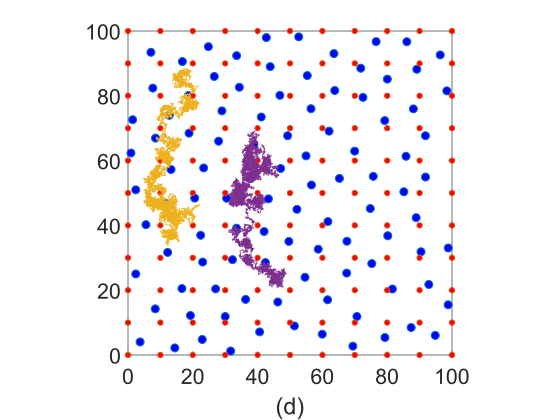}
\end{tabular}
\caption{\small  The real space configuration of the vortex lattice at various temperatures and drives. The small red discs are the
pinned, immobile vortices and the blue, larger discs are the interstitial vortices. In (b), (c), and (d), the trajectories of two of the vortices
have been plotted for $50,000$ time steps. (a) The ground state structure obtained for temperature $T = 0.001$ and for zero drive. (b) A snap 
shot of the low temperature ($T = 0.001$) vortex lattice with drive, $F_d = 86 F_b$. Note that the trajectories of the two vortices shown
are very narrow as the lateral excursions are minimal. (c) Lattice at a given instant for the same drive as above but at a temperature just below the
melting temperature, $T_c = 0.7$. There is now a substantial lateral (in the direction perpendicular to the drive) vibrations in the lattice. But the particles
are not yet freely diffusing. (d) Snapshot of the lattice, again for the same driving force but at a higher temperature ($T = 1.4$) at which the lattice has 
melted. Note that the vortices are now diffusing laterally and their motion is largely independent other than the average drift under the driving force.}
\label{Fig09}
\end{figure}

To understand what is happening to the vortex lattice itself as it is driven, we studied the structure of the vortex system 
(expressed in terms of the structure factors at $\vec G_1$ and $\vec G_2$ ) in the same temperature range. The structure factors
were computed numerically by evaluating $S(\vec G_i) = \frac{1}{N^2}<\sum_{m,n} e^{j \vec G_i \cdot (\vec r_m - \vec r_n)}>$ ($i = 1, 2$), 
where $N$ is the number of vortices in the system and $\vec r_m$ and $\vec r_n$ are the positions of the $m$th and the $n$th vortices. 
The results shown in Fig.~\ref{Fig04} is very similar to the case when there is no drive \cite{toby1} and shows the presence of a 
continuous transition close to $T_c = 0.7$. One is thus able to correlate the minimum of the vortex current as a function of temperature to the melting 
of the vortex lattice. In Fig.~\ref{Fig09} the real space configuration of the lattice for various temperatures are shown. To get a glimpse of the individual vortex
dynamics, trajectories of two vortices for $50,000$ time steps during the simulation are also shown. 
In the absence of drive, the low temperature ($T = 0.001$) phase of the interstitial vortices (blue, larger discs in the figure) is the partially pinned 
solid (Fig.~\ref{Fig09}(a)). When the lattice is driven above the depinning threshold, it drifts with respect to the substrate. 
(Fig.~\ref{Fig09}(b)) shows the snapshot of the driven lattice (driven along the easy direction) at $T = 0.001$ for driving force $F_d = 86 F_b$. 
The lattice lattice moves as a whole and the trajectory of two of the vortices show that the motion is confined to a narrow channel. Moreover they are
drifting in tandem, as expected. As the temperature is brought close to the melting temperature (see (Fig.~\ref{Fig09}(c)), the lateral motion of the vortices 
in the lattice is enhanced but the vortices are still confined to the vertical channels. For temperatures well above $T_c$, the vortices are disordered and are 
moving independently with a net drift due to the applied drive (Fig.~\ref{Fig09}(d)).

This phenomenon looks similar to the {\it peak effect} seen in vortex systems with random pinning, where the critical 
depinning current $J_c$ peaks as the vortex lattice melts. To see this effect unambiguously in the current system, one has to keep the driving force close to the 
threshold value and vary the temperature. Simulations near the depinning force have not been carried out due to the fact that thermal noise is 
dominant at such low drives and the effects are less discernible for the system size and number of time steps we have used. This is more so 
when one is close to the transition temperature which is a second order one as seen from the equilibrium simulations. The minimum in vortex
current close to the transition temperature should lead to a corresponding increase in the depinning force when one nears the transition 
temperature.

\subsection{Explanation of the observed minima in vortex current - temperature curve}
One would have expected the vortex current to increase as the temperature is increased with the drive force held constant, 
since thermal activation would help in crossing the potential barriers that the vortices encounter as they move across the sample
\cite{koshelev}. The fact that this simple picture does not lead to the observed behavior in the current system indicates the presence 
of some mechanism which impedes vortex motion as the transition temperature is approached from below.
A possible explanation for the phenomenon observed is that the effective strength of barriers encountered by the vortices may increase with increasing temperature 
in the solid phase. The rigidity of the lattice in the solid phase means that the vortices cannot individually move across the barriers. On the other hand
the increase in thermal vibrations in the lateral direction (that is, direction perpendicular to the drive direction) as temperature is increased (see vortex trajectories in 
(Fig.~\ref{Fig09} (c)) can on the average increase the energy barriers encountered by the vortices in the lattice. Below, we develop this idea in to a semi quantitative 
analysis and argue that this is a possible explanation for the drop in current observed. 

As the partially pinned vortex solid is driven in the easy direction, the drive direction ensures that at low temperatures it passes through relatively easy paths, 
i.e. paths along which the potential produced by the pinning centers has relatively low and flat maxima. This is so because at low 
temperatures (much below the transition temperature), typical paths taken by the vortices are close to lines that bisect the sides of 
the pin squares (see vortex trajectories in Fig.~\ref{Fig09} (b)) . In a coordinate system where the drive is in the $y$ direction and the origin is at the corner of a square unit cell of 
the pin lattice, such paths are lines parallel to the $y$ axis passing through $x=\pm d/2, \pm 3d/2,...$ etc (paths like the one
along $pq$ in Fig.~\ref{Fig01} (B)). Paths parallel to $y$ direction but away from the this bisector lines will have larger barriers to be crossed (paths like the one
along $rs$ in Fig.~\ref{Fig01} (B)).   
The inset of Fig.~\ref{Fig06} shows the height of the maximum of the pinning potential along paths parallel to the $y$ axis as a function of the 
lateral coordinate, $x$ (that is along the line $ab$ in Fig.~\ref{Fig01} (B)). This is given by the function
\begin{equation} 
U(x, y = 0) = U_0 \Big[K_0(\frac{x}{\lambda}) + K_0(\frac{x-d}{\lambda})\Big]\;,
\end{equation}
where $U_0 = A F_0 \lambda_{eff}$ and $K_0$ is the modified Bessel function of the second kind. $U(x,y)$ is the potential corresponding to the
substrate and for finding potential along the edge of the square we have only retained the contribution to the substrate potential 
from the closest two pinned vortices at $(0,0)$ and $(0,d)$. The path that goes through the centers of the pin squares is clearly the one with the 
smallest value of the pinning potential barrier. As the temperature approaches the transition point from below, the spread of the average particle density in the transverse 
direction ($x-$ direction in Fig.~\ref{Fig01} (A)) increases. 
Since the maximum pinning potential encountered along a path increases as the path moves away from the center of the pin squares, 
an increase in the spread of the density distribution in the $x$ direction could lead to the lattice encountering higher barriers on the average. Below we make a quantitative 
estimate of the average barrier heights as a function of temperature to see whether there is a sudden spike in its value as the transition point is approached.
\begin{figure}[htbp]
\center
\includegraphics[height=7cm]{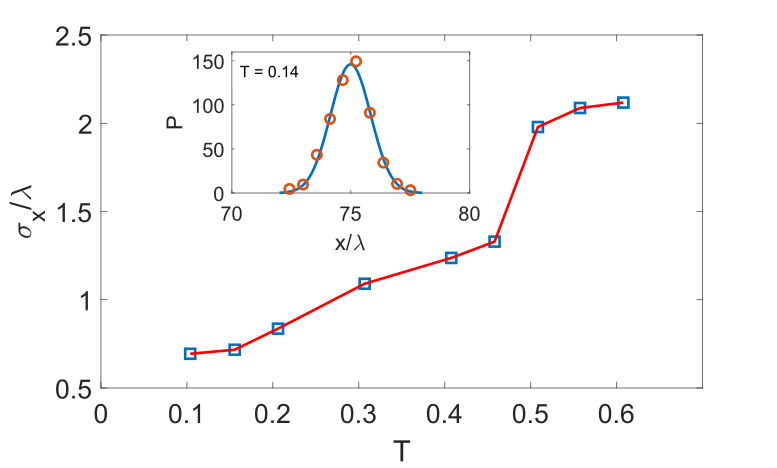}
\vspace{0.0cm}
\caption{\small The main plot shows the width of distribution of particle position in the $x$ direction as a function of temperature. There is a
gradual increase in the width capturing the increased thermal vibration of the vortices. The inset shows the distribution of the $x$ coordinate of a 
single vortex at $T = 0.14$. The width of the Gaussian is used to find the average barrier height that the driven vortices encounter (Fig.~\ref{Fig06}). 
Note that the Gaussian in the inset is not normalized.}
\label{Fig05}
\end{figure}

Consider the vibration of particles in the direction perpendicular to the drive (the $x$ direction). As the temperature goes up, the distribution of the 
$x$ coordinate gets broader. One can estimate the amount of increased barrier height that the vortices encounter due 
to this spread by finding the width of the distribution as a function of temperature and averaging the peak pinning potential as a function of $x$ 
over such distributions. The width of the distribution of the $x$-coordinate of the vortices as a function of 
temperature, $\sigma_x (T)$, was determined from the zero drive simulations by finding the room mean square displacement (rms) of
the vortices in the $x$ direction:
\begin{equation}
\sigma_x(T) = \sqrt{<\frac{1}{N}\sum_i (x_i -  \bar x_{i})^2>}
\end{equation}
where the sum is over the $N$ vortices, $\bar x_{i}$ is the mean position of the $i$th vortex and angular brackets represents average over the
Brownian dynamics moves in equilibrium. One has to be careful while calculating the width, since the particles sometimes diffuse in the $x$ direction 
to a nearby pinning square. Computing the rms displacement without due care would lead to an erroneous estimate. The result obtained by keeping track of the 
individual particle motion and computing the width of vibrations within a pinning square is shown in Fig.~\ref{Fig05}. As expected there is a gradual
increase in the amount of excursion that each vortex makes in the $x$-directions as temperature goes up. Note that beyond the melting temperature
the above quantity is ill defined as the vortices have melted and does not have a well defined mean position. 

We perform a uniform average of the potential barrier, $U(x,0)-U_b$ over the width $\sigma_x(T)$ to obtain an estimate of the effective 
barrier height, $\Delta E_b$, at different temperatures. Here $U_b = 4 U_0 K_0(5 \sqrt(2)$ is the substrate potential energy at the 
center of the pinning square. 
That is,
\begin{equation}
\Delta E_b(T) = \frac{1}{2 \sigma_x(T)} \int_{d/2 - \sigma_x}^{d/2 + \sigma_x} (U(x,0) - U_b) dx
\end{equation}
The variation of the average barrier height with temperature is shown in Fig.~\ref{Fig06}. The effective barrier height exhibits a  sudden 
increase just below $T_c$. The sharp rice in the barrier height can be attributed to the increase in $\sigma_x$ with temperature coupled
to the fact that $U(x,0)$ rises sharply as $x$ moves closer to values $0$ and $x = d$. This effective increase in barrier height is a 
possible explanation for the {\it peak effect} like feature seen in our simulations. The average barrier height that the vortex lattice encounters goes 
up with temperature leading to an increased resistance to flow. In the regime beyond the melting temperature, individual vortex motion are largely 
uncorrelated. With the rigidity of lattice no more a constraining factor, the individual vortex drift along the driving force direction is aided by the 
increasing temperature.  
\begin{figure}[htbp]
\center
\includegraphics[height=7cm]{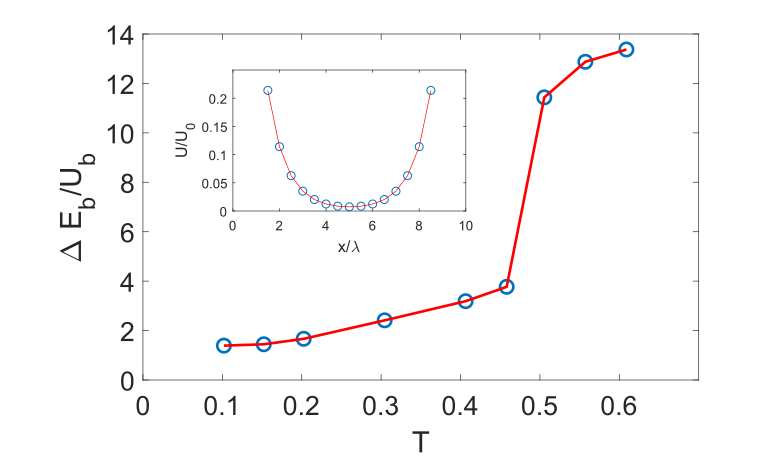}
\vspace{0.0cm}
\caption{\small The main plot shows the effective barrier height for a vortex as a function of temperature. Notice the sharp rise in
the barrier height as the transition temperature is approached. The inset shows the substrate potential value along a side of the pin
square (say, along line $ab$ in Fig.~\ref{Fig01}(B)). This gives the maximum pinning potential that a vortex has to cross as it tries to 
move from one pinning cell to the next. The average energy barrier is given in units of $U_b$, which is the minimum barrier height (see text).}
\label{Fig06}
\end{figure}

The arguments based on an increase in the effective barrier height due to increased thermal wandering of the vortices in the transverse 
direction provide a semi quantitative explanation of the observed temperature dependence of the vortex current. These arguments look definitely 
plausible for describing the behavior for driving force that are slightly higher than the depinning force. For such values of
the driving force, vortices that do not move along the lowest-energy paths (i.e.paths passing through the centers of the pin squares)
would encounter potential barriers even when the force on them due to the driving force is taken into account. However, for drive strengths much 
higher than the depinning force (most of our simulations were done in this regime), these arguments need more justification.  We believe that our 
arguments would remain qualitatively valid even for strong drives. This is because the speed of a vortex moving along a path parallel to the $y$ axis would still depend 
on the variation of the pinning potential along the path, because the force on the vortex has a contribution arising from the pinning potential. 
So, vortices moving along paths for which the pinning potential has high and steep maxima (such as paths that are substantially displaced 
from the centers of the pin squares) would be slowed down in comparison to those moving in paths along which the peaks of the pining potential 
are low (such as paths lying near the centers of of the pin squares). Since the typical transverse displacement of the vortices increases with 
increasing $T$, the probability of a vortex taking a path that is substantially displaced from the centers of the squares would increase as the 
temperature is increased. Since vortices move more slowly along such paths, this may lead to a decrease in the vortex current.

\begin{figure}[htbp]
\center
\includegraphics[height=7cm]{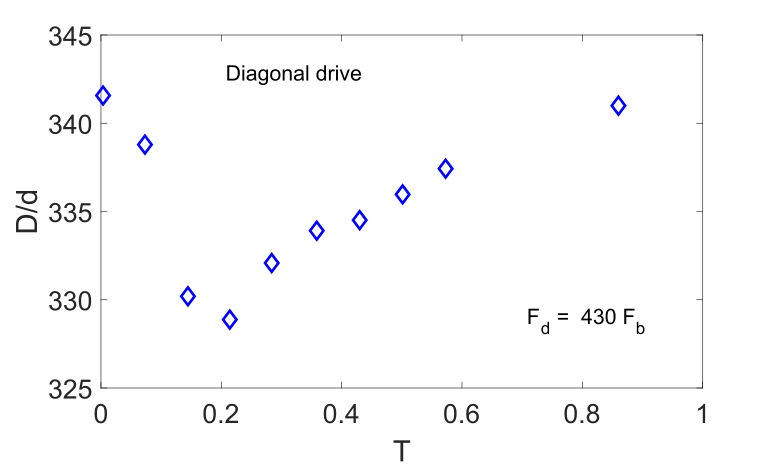}
\vspace{0.0cm}
\caption{\small  Plot showing variations of average particle velocity with temperature for driving force along the diagonal direction. The value of the 
substrate strength parameter, $A = 0.0001$. The average displacement of the vortices, $D$, for a fixed time ($5 \times 10^5$ time steps) is 
plotted as a measure of the average vortex velocity. The driving force applied is $F_d = 430 F_{b}$. Like for the case of easy direction drive, 
there is a minimum located at the vicinity of the melting transition to the modulated liquid from the partially pinned solid phase (see Fig.~\ref{Fig08}).}
\label{Fig07}
\end{figure}
\subsection{Driven lattice along the diagonal direction}
\begin{figure}[htbp]
\center
\includegraphics[height=7cm]{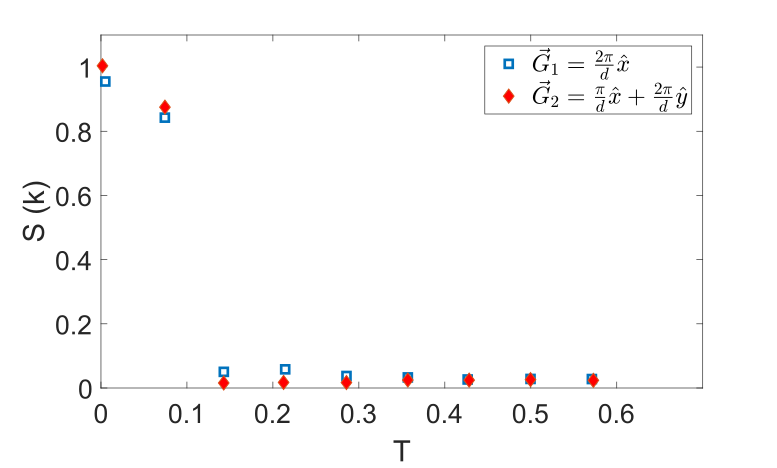}
\vspace{0.0cm}
\caption{\small  Structure factor at the two smallest reciprocal lattice vectors as a function of temperature. The value of the 
substrate strength parameter, $A = 0.0001$. The driving force applied is $F_d =  430 F_{b}$ the same as the one used for
finding the average vortex velocity (see Fig.~\ref{Fig07}). There is a transition to modulated liquid at temperature close to $T = 0.14$.}
\label{Fig08}
\end{figure}

In order to check the robustness of the phenomenon seen above for the easy direction drive case, we have also studied how the vortex
drift varied as a function of temperature when the lattice is driven along the direction diagonal to the pinning square.  
Fig.~\ref{Fig07} shows the vortex current as a function of temperature for driving force magnitude $F_d = 430 F_b$. The corresponding behavior of structure factor
corresponding to reciprocal lattice vectors $\vec G_1$ and $\vec G_2$ are shown in Fig.~\ref{Fig08}. The structure factor plot indicates a melting transition close to
$T = 0.14$ in this case. Unlike the easy direction drive, the melting of the lattice in this case takes place at a lower value of temperature compared 
to the undriven case. Moreover, the sharp drop in structure factor values indicates that the transition is more abrupt, possibly of a first order kind.
The vortex current versus temperature plot shows that close to the transition, there is a dip in vortex current just as in the previous case we studied.
This clearly indicates that the resistance to motion of the driven lattice with respect to the substrate is intimately related to the melting transition.
It is interesting to note that for both the drive directions the ordering corresponding to the reciprocal lattice vector $\vec G_2$ drops to zero faster
than that at $\vec G_1$. And the location of minima of the vortex current seems to be correlated to the ordering at $\vec G_1$. This ties in well with our
explanation that it is the disorder in the lateral direction that is leading to larger resistance to vortex flow.

\section{Summary} We have carried out Brownian dynamics simulation studies on driven partially pinned solid in two dimensions 
formed in the presence of a substrate with square symmetry. The lattice undergoes elastic depinning as the
drive force is increased at low temperatures. For large drives we observed a dip in vortex velocity as the transition temperature is approached from
below, much like the {\it peak effect} seen in systems with random disorder. We have attributed this phenomenon to an effective
increase in the potential barrier that the vortices encounter close to the transition temperature. This argument is not unrelated to the explanation of the 
{\it peak effect} in terms of a softening of the vortex lattice as the melting transition is approached from below. The increase 
in the width parameter $\sigma_x$ near $T= 0.4$ (see Fig.~\ref{Fig05}) corresponds to a pre-transition softening of the vortex lattice leading to
a more efficient pinning as seen by the corresponding rise in $\Delta E_b$. Tough the barrier analysis was done
only for the case of easy direction drive, our data shows that the dip in vortex velocity happens for drives along other 
directions too. A similar analysis for these cases are more involved due to the reduced symmetry of these drive directions.

A related system and process has been simulated in the work by Reichhardt et. al. \cite{reichhardt6} where a single colloidal particle
was dragged through a two dimensional colloidal crystal lattice and the drag force on the drag particle was studied.
They have seen an increase in the drag force as the melting temperature is approached and have attributed it to local 
melting induced by the dragged particle. The present study is different in two key ways. In the present case, the lattice is
driven against a extended substrate and the substrate is fixed unlike the drag particle in the earlier work. It is interesting to note that
in both system one ends up with similar effects. We have not carried out a defect based analysis of the crystal structure as was
done in that work. We propose to carry out this study in the future.

Though the simulations have been carried out for a specific kind of interaction and substrate potential relevant for vortex motion
in decorated thin films, the physics should be accessible in other systems too. In colloidal systems, where there is
a greater control over the system parameters and better experimental accessibility to the microscopic details, study of these
phenomena will be interesting. On the simulation front, it would be interesting to look at the depinning phenomenon itself as a 
function of temperature. This would require larger systems and extended time averages as the thermal noise can make
identification of the depinning of the lattice difficult. From the current study, one can anticipate a non trivial behavior of
the threshold depinning force as the temperature is varied. One can also address questions on the nature of the elastic depinning
transition vis-\'a-vis its critical properties and how the nature of depinning varies with filling factors not equal to but close to one.
Another aspect that has to be explored further is the nature of the transitions between the moving crystal and fluid phase in the
presence of the drive. As seen above, the drive direction can alter appreciably the transition point and the nature of transition. Further
study is required to characterize the nature of these transitions. Studies are on to address these questions.

\section{Acknowledgment}
The author would like to acknowledge Prof. C. Dasgupta, P. Shatrhunjay and C. Jadeja  for useful discussions.
This is a pre-print of an article published in Physica A: Statistical Mechanics and Applications. The final authenticated version is available online 
at: https://doi.org/10.1016/j.physa.2020.124737



%


\end{document}